\documentstyle[preprint,aps]{revtex}

\begin{document}
\title{Mechanisms of spin-polarized current-driven magnetization switching }
\author{S. Zhang}
\address{Department of Physics and Astronomy, University of\\
Missouri-Columbia, Columbia, MO 65211\\
P. M. Levy \\
Department of Physics, New York University, 4 Washington Place, \\
New York, NY 10003 \\
A. Fert \\
Unit{\'{e}} Mixte de Physique CNRS/THALES (CNRS-UMR 137) \\
Domaine de Corbeville, 91404 Orsay, France}
\maketitle

\begin{abstract}
The mechanisms of the magnetization switching of magnetic multilayers driven
by a current are studied by including exchange interaction between local
moments and spin accumulation of conduction electrons. It is found that this
exchange interaction leads to two additional terms in the
Landau-Lifshitz-Gilbert equation: an effective field and a spin torque. Both
terms are proportional to the transverse spin accumulation and have
comparable magnitudes.
\end{abstract}

\date{\today}


\narrowtext

The concept of switching the orientation of a magnetic layer of a
multilayered structure by the current perpendicular to the layers was
introduced by Slonczewski\cite{slon} and Berger \cite{berger}, and has been
followed up by Waintal et al.\cite{waintal}. The central idea is that for a
noncollinear configuration of the moments of the magnetic layer the current
induces a torque acting on the spins of the conduction electrons which in
turn transmit this torque to the background magnetization of the magnetic
layers through the exchange interaction between conduction electrons and the
local ``d'' electrons. An alternate mechanism of current induced switching
was put forth by Heide et al \cite{heide} in which the current across the
magnetically inhomogeneous multilayer produces spin accumulation which
establishes an energy preference for a parallel or antiparallel alignment of
the moments of the magnetic layers; this magnetic ``coupling'' was posited
to produce switching. Recent experiments have reliably demonstrated that the
magnetization of a magnetic layered structure is indeed switched back and
forth by an applied current \cite{Katine,Albert,Fert}. However, it is
unclear whether the magnetization switching is triggered by the
current-driven effective field or by the spin torque mechanism or both.

Here we examine the two views of current induced switching, spin torque and
effective field, by solving the equations of motion for the spin
accumulation and the local magnetization. We find the two mechanisms do
coexist; albeit in form very different from that envisaged by the above
referenced authors. The salient difference between our treatment of spin
diffusion and previous treatments \cite{Son,Johnson,valet}, lies in the
inclusion of the exchange interaction between the spin accumulation and the
magnetic background. With our results, we can understand these two
mechanisms on an equal footing: both are simultaneously derived and both
depend on the same set of parameters used for understanding the giant
magnetoresistance when the current is perpendicular to the plane of the
layers (CPP). Furthermore, we have introduced a new length scale for the
transverse spin accumulation and clarified the ferromagnetic layer thickness
dependence of the switching dynamics.

Let us consider a magnetic multilayer with the current perpendicular to the
plane of the layer (defined as $x$-direction). The linear response of the
current to the electrical field can be written as a spinor form,
\begin{equation}
\hat{\jmath}(x)=\hat{C}E(x)-\hat{D}\frac{\partial \hat{n}}{\partial x},
\end{equation}
where $E(x)$ is the electric field, $\hat{\jmath}$, $\hat{C}$, $\hat{D}$,
and $\hat{n}$ are the $2\times 2$ matrices representing the current, the
conductivity, the diffusion constant, and the {\em accumulation} at a given
position. The diffusion constant and the conductivity are related via the
Einstein relation $\hat{C}=e^{2}\hat{N}(\epsilon _{F})\hat{D}$ for a
degenerate metal, where $\hat{N}(\epsilon _{F})$ is the density of states at
the Fermi level. In general, one can express these matrices in terms of the
Pauli spin matrix $\mbox{\boldmath $\sigma$}$,
\begin{equation}
\hat{C}=C_{0}\hat{I}+\mbox{\boldmath $\sigma$}\cdot {\bf C},
\end{equation}
\begin{equation}
\hat{D}=D_{0}\hat{I}+\mbox{\boldmath $\sigma$}\cdot {\bf D},
\end{equation}
and
\begin{equation}
\hat{n}=n_{0}\hat{I}+\mbox{\boldmath $\sigma$}\cdot {\bf m},
\end{equation}
where $2n_{0}$ is the charge accumulation and ${\bf m}$ is the spin
accumulation. By placing Eqs.~(2)-(4) into (1), we rewrite the linear
response in terms of the electric current $j_{e}$ and magnetization current $%
{\bf j}_{m}$ as
\begin{equation}
j_{e}\equiv {\rm Re}({\rm Tr}\hat{\jmath})=2C_{0}E(x)-2D_{0}\frac{\partial
n_{0}}{\partial x}-2{\bf D}\cdot \frac{\partial {\bf m}}{\partial x},
\end{equation}
and
\begin{equation}
{\bf j}_{m}={\rm Re}{\rm Tr}(\mbox{\boldmath $\sigma$}\hat{\jmath})=2{\bf C}%
E(x)-2{\bf D}\frac{\partial n_{0}}{\partial x}-2D_{0}\frac{\partial {\bf m}}{%
\partial x}.
\end{equation}
It is noted that we have chosen the unit $e=\mu _{B}=1$ for the notation
convenience so that the electrical current and the magnetization current
have the same unit. For a transition ferromagnet, one defines the spin
polarization parameter $\beta $ as ${\bf C}=\beta C_{0}{\bf M}_{d}$, where $%
{\bf M}_{d}$ is the {\em unit} vector to represent the direction of the
local magnetization. Similarly, we can introduce a spin polarization $\beta
^{\prime }$ for the diffusion constant ${\bf D}=\beta ^{\prime }D_{0}{\bf M}%
_{d}$. These two polarization parameters are not necessarily the same, i.e.,
when the density of states are different for spin up and down
electrons, $\beta \neq \beta ^{\prime }$. By inserting these relations into
Eqs.~(5) and (6), and eliminating the electric field and charge density, we
obtain
\begin{equation}
{{\bf j}_{m}}=\beta j_{e}{\bf M}_{d}-2D_{0}\left[ \frac{\partial {\bf m}}{%
\partial x}-\beta \beta ^{\prime }{\bf M}_{d}({\bf M}_{d}\cdot \frac{%
\partial {\bf m}}{\partial x})\right] ,
\end{equation}
where we have dropped an uninteresting term proportional to the derivative
of the charge accumulation $\partial {n_{0}}/\partial x.$ This result is
similar to that obtained by Heide \cite{heide2}.

We now describe the equations of motion for the spin accumulation and local
magnetization when we turn on the interaction between the spin accumulation
and the local moment via the sd contact interaction,
\begin{equation}
H_{int}=-J{\bf m}\cdot {\bf M}_{d}.
\end{equation}
With this interaction, the equation of motion for the spin accumulation is
\begin{equation}
\frac{d{\bf m}}{dt}+(J/\hbar ){\bf m}\times {\bf M}_{d} =-\frac{{\bf m}}{%
\tau _{sf}},
\end{equation}
where $\tau _{sf}$ is the spin-flip relaxation time of the conduction
electron. The second term on the left hand side represents the processional
motion of the accumulation due to the sd interaction when the magnetization
directions of the spin accumulation and the local moments are not parallel.
Since the conduction electrons carry a spin current given by Eq.~(7), we
replace $\frac{d{\bf m}}{dt}$ by $\frac{\partial {\bf m}}{\partial t}+\frac{{%
\partial {\bf j}}_{m}}{\partial x}$. By using Eq.~(7), we find
\begin{equation}
\frac{1}{2D_{0}}\frac{\partial {\bf m}}{\partial t}=\frac{\partial ^{2} {\bf %
m}}{\partial x^{2}}-\beta \beta^{\prime}{\bf M}_{d}\left( {\bf M}_{d} \cdot
\frac{\partial ^{2}{\bf m}}{\partial x^2}\right) -\frac{{\bf m}}{\lambda
_{sf}^{2}}-\frac{{\bf m} \times {\bf M}_{d}}{\lambda _{J}^{2}} ,
\end{equation}
where we have defined $\lambda _{sf}\equiv \sqrt{2D_{0}\tau _{sf}}$ and $%
\lambda _{J}\equiv \sqrt{2\hbar D_{0}/J}$ \cite{lambdaJ}. The latter gives
rise to a new length scale which governs the spin torque created by the
current. The significance of this new length scale will be discussed later.

The equation of motion for the local magnetization is
\begin{equation}
\frac{d{\bf M}_{d}}{dt}= - \gamma _{0}{\bf M}_{d} \times ({\bf H}_{e}+J {\bf %
m}) +\alpha {\bf M}_{d}\times \frac{d{\bf M}_{d}}{dt},
\end{equation}
where $\gamma_0$ is the gyromagnetic ratio, ${\bf H}_{e}$ is the magnetic
field including the contributions from the external field, anisotropy and
magnetostatic field, the additional effective field $J{\bf m}$ is due to
coupling between the local moment and the spin accumulation, and the last
term is the Gilbert damping term.

To solve for the dynamics of the spin accumulation and the local moment, we
need to simultaneously determine them using Eqs.~(10) and (11). The 
time scales are very different for the spin
accumulation and the local moments. The characteristic time scales of the
former are of the order of $\tau _{sf}$ and $\hbar /J$, see Eq.~(9),
i.e., of the order of picoseconds ($10^{-12}$ seconds). For the local moment, 
the time scale is $\gamma_0^{-1} (H_e + Jm_{\perp})^{-1}$. For a magnetic
field of the order of 0.1 Tesla, this time scale is of the order of
nanoseconds. Therefore, as long as one is
interested in the magnetization process of the local moments, one can always
treat the spin accumulation in the limit of long times. The two dynamic
equations are then simply decoupled: we first solves Eq.~(10) with fixed
local moments (independent of time) and set the left hand side of Eq.~(10)
to zero. Once the spin accumulation is obtained, we substitute it into
Eq.~(11) to solve the dynamics of the local moments.

Before we proceed to solve for the {\em stationary }solution of Eq.~(10),
let us first discuss the general features derived from Eq.~(10). We separate
the spin accumulation into longitudinal (parallel to the local moment) and
transverse (perpendicular to the local moment) modes. Equation (10) can now
be written as
\begin{equation}
\frac{\partial ^{2}{\bf m}_{||}}{\partial x^{2}}-\frac{{\bf m}_{||}}{\lambda
_{sdl}^{2}}=0,
\end{equation}
where $\lambda _{sdl}=\sqrt{1-\beta \beta ^{\prime }}\lambda _{sf}$, and
\begin{equation}
\frac{\partial ^{2}{\bf m}_{\perp }}{\partial x^{2}}-\frac{{\bf m}_{\perp }}{%
\lambda _{sf}^{2}}-\frac{{\bf m}_{\perp }\times {\bf M}_{d}}{\lambda _{J}^{2}%
}=0.
\end{equation}
The longitudinal spin accumulation ${\bf m}_{||}$ decays at the length scale
of the spin diffusion length $\lambda _{sdl}$ while the transverse spin
accumulation ${\bf m}_{\perp }$ decays as $\lambda _{J}$. The spin diffusion
length $\lambda _{sdl}$ has been measured to be about 60nm in Co \cite{Bass}. 
We estimate $\lambda _{J}$ by taking the typical diffusion constant of a
metal to be $10^{-3}$ (m$^{2}$/s) and $J=0.1-0.4$ (eV) so that $\lambda _{J}$
is about 1.2 nm to 2.4 nm. Thus, the transverse spin accumulation has a much
shorter length scale compared to the longitudinal one.

Before we apply Eqs.~(12) and (13) to a multilayer structure, we take a look
at the influence of the spin accumulation on the local moment. As seen from
Eq.~(11), the longitudinal spin accumulation has no effect on the local
moment. We may re-write Eq.~(11) in terms of the transverse spin
accumulation only,
\begin{equation}
\frac{d{\bf M}_{d}}{dt}=-\gamma _{0}{\bf M}_{d}\times ({\bf H}_{e}+J{\bf m}%
_{\perp })+\alpha {\bf M}_{d}\times \frac{d{\bf M}_{d}}{dt}.
\end{equation}
To discuss the transverse accumulation we introduce a vector $A$
such that $J{\bf m}_{\perp }={\bf A} \times {\bf M}_{d}$. 
If one considers a system with
two ferromagnetic layers whose magnetization directions are not parallel to
each other, the spin accumulation at one layer depends on the orientation of
the other. Let us suppose that the above equation is used for the layer F1,
i.e., denote ${\bf M}_{d}={\bf M}_{d}^{(1)}$. 
The magnetization of the other layer
is labeled as ${\bf M}_{d}^{(2)}.$ Without loss of generality, we can write
the two components of the
accumulation in the plane transverse to ${\bf M}_{d}^{(1)}$ as
$J{\bf m}_{\perp }=a{\bf M}_{d}^{(2)}\times
{\bf M}_{d}^{(1)}+b({\bf M}_d^{(1)} \times {\bf M}_{d}^{(2)} ) \times 
{\bf M}_d^{(1)}$, where $a$ and $b$ are determined by geometric
details of the multilayer. Placing this into Eq.~(14), we find 
\begin{equation}
\frac{d{\bf M}_{d}^{(1)}}{dt}=-\gamma _{0}{\bf M}_{d}^{(1)}\times ({\bf H}%
_{e}+b{\bf M}_{d}^{(2)})-\gamma _{0}a{\bf M}_{d}^{(1)}\times ({\bf M}%
_{d}^{(2)}\times {\bf M}_{d}^{(1)})+\alpha {\bf M}_{d}^{(1)}\times \frac{d%
{\bf M}_{d}^{(1)}}{dt}.
\end{equation}
Thus the transverse spin accumulation produces two effects
simultaneously ( one can call them either fields or torques): one is $%
b{\bf M}_{d}^{(2)}$ the ``effective field'' which gives rise to a
precessional motion and the other is $a{\bf M}_{d}^{(1)}\times 
({\bf M}_{d}^{(2)}\times {\bf M}_{d}^{(1)})$ which is 
called  the ``spin torque''. Both terms lead to
significant corrections to the original Landau-Lifshitz-Gilbert equation. It
has been shown that both terms are capable to switch the magnetic moments
\cite{koch}. Note the effective field introduced here looks as if it arises
from the current induced coupling named NEXI, however it is different as
NEXI was attributed to the longitudinal component of the spin accumulation
\cite{heide2}. In contrast we have shown that only the transverse spin
accumulation must be taken into account and that the longitudinal
accumulation does not produce any effect on the motion of local moments. 
An even more striking difference is Heide's finding that 
``the presence of a second ferromagnetic layer is not necessary''. 
This is because his longitudinal accumulation exists for a single
F layer, while a second F layer with tilted magnetization is required for
transverse accumulation and for our mechanism. It
is notable that the ``torque'' term, first introduced by J. Slonczewski,
appears on an equal footing with the effective field $b{\bf M}_{d}^{(2)}$ as
both are related to the {\em transverse} spin accumulation.

We now explicitly verify that the solution of the transverse accumulation $%
{\bf m}_{\perp }$ indeed has our proposed general form and we quantitatively
determine the magnitude of the effective field (proportional to $b$ term)
and the `` spin torque'' (proportional to $a$ term) entering Eq.~(15). To
obtain a physically transparent solution of the spin accumulation, we choose
an oversimplified case to perform our calculation so that the effective
field and spin torque can be analytically derived. We consider a system
consisting of a very thick ferromagnetic layer which is assumed to be
pinned, a spacer layer which is infinitely thin so that the spin current is
conserved across the layer when there is no spin flip scattering in this
region, and a thin ferromagnetic layer backed by an ideal paramagnetic
layer. In addition we make our calculation simpler by neglecting
spin-dependent reflection at the interfaces. In such a system, we look for
the solution of the spin accumulation in the thin F1 layer whose
magnetization direction is at the positive $z$-direction. The magnetization
direction of the pinned layer is ${\bf M}_{d}^{(2)}=\cos \theta {\bf e}%
_{z}-\sin \theta {\bf e}_{y}$ where $\theta $ is the angle between ${\bf M}%
_{d}^{(2)}$ and ${\bf M}_{d}^{(1)}={\bf e}_{z}$. From Eqs.~(12) and (13),
and by assuming the same $\lambda _{sdl}$ for the thin magnetic layer, F1,
as for the non-magnetic layer which backs it, we write the solution for the
F1 layer as
\begin{equation}
m_{z}(x)=G_{1}\exp {(-x/\lambda }_{sdl}{)}
\end{equation}
\begin{equation}
m_{x}(x)=G_{2}\exp {(-x/l_{+})}+G_{3}\exp {(-x/l_{-})}
\end{equation}
\begin{equation}
m_{y}(x)=-iG_{2}\exp {(-x/l_{+})}+iG_{3}\exp {(-x/l_{-})}
\end{equation}
where $l_{\mp }^{-1}=\sqrt{\frac{1}{\lambda _{sf}^{2}}\pm \frac{i}{\lambda
_{J}^{2}}}$. To determine the constants of integration, we assume the thick
magnetic layer F2 is half metallic so that the current is fully spin
polarized and we demand that the spin current is continuous across $F2/N/F1$
interface \cite{footnote}; we find
\begin{equation}
\beta j_{e}-2D_{0}(1-\beta \beta ^{\prime })\left( -\frac{G_{1}}{\lambda
_{sdl}}\right) =j_{e}\cos \theta ,
\end{equation}
\begin{equation}
-2D_{0}\left( \frac{G_{2}}{l_{+}}+\frac{G_{3}}{l_{-}}\right) =0,
\end{equation}
and
\begin{equation}
-2D_{0}(-i)\left( -\frac{G_{2}}{l_{+}}+\frac{G_{3}}{l_{-}}\right)
=-j_{e}\sin \theta .
\end{equation}
Thus we determine the constants to be
\begin{equation}
G_{1}=-\frac{j_{e}\lambda _{sdl}(\beta -\cos \theta )}{2D_{0}(1-\beta \beta
^{\prime })},
\end{equation}
\begin{equation}
G_{2}=\frac{j_{e}l_{+}\sin \theta }{4iD_{0}},
\end{equation}
and
\begin{equation}
G_{3}=-\frac{j_{e}l_{-}\sin \theta }{4iD_{0}}.
\end{equation}
Therefore, we find the transverse spin accumulation
\begin{equation}
{\bf m}_{\perp }=-\left( \frac{j_{e}}{2D_{0}}\right) \left[ {\rm Im}%
(l_{+}e^{-x/l_{+}}){\bf M}_{d}^{(2)}+{\rm Re}(l_{+}e^{-x/l_{+}}){\bf M}%
_{d}^{(2)}\times {\bf M}_{d}^{(1)})\right] \times {\bf M}_{d}^{(1)}
\end{equation}
where we have used $-\sin \theta {\bf e}_{x}={\bf M}_{d}^{(2)}\times {\bf M}%
_{d}^{(1)}$ and $\sin \theta {\bf e}_{y}=({\bf M}_{d}^{(2)}\times {\bf M}%
_{d}^{(1)})\times {\bf M}_{d}^{(1)}$. We immediately see that the form of
the spin accumulation given above is precisely the form we used in deriving
Eq.~(15). To obtain the coefficients $a$ and $b$ entering Eq.~(15) we
average this spin accumulation over $0\leq x\leq t_{F}$ where $t_{F}$ is the
thickness of the F1 layer and find
\begin{equation}
a=-\frac{Jj_{e}}{2D_{0}t_{F}}{\rm Im}[l_{+}^{2}(1-e^{-t_{F}/l_{+}})]
\end{equation}
and
\begin{equation}
b=\frac{Jj_{e}}{2D_{0}t_{F}}{\rm Re}[l_{+}^{2}(1-e^{-t_{F}/l_{+}})],
\end{equation}

It is noted that both $a$ and $b$ change sign under time reversal.
The former agrees with that found in \cite{slon,berger,waintal}; 
while the latter has not been considered by these authors.
To estimate $a$ and $b$, we take the limit $\lambda _{sf}\gg \lambda _{J}$
in which case $l_{+}=(1+i)\lambda _{J}/\sqrt{2}$. By placing this into
Eqs.~(26) and (27), we find
\begin{equation}
a= - \frac{\hbar j_{e}a_{0}^{3}}{\sqrt{2}e\mu _{B}\lambda _{J}}\left( \frac{%
1-\cos \xi e^{-\xi }}{\xi }\right)
\end{equation}
and
\begin{equation}
b= \frac{\hbar j_{e}a_{0}^{3}}{\sqrt{2}e\mu _{B}\lambda _{J}}\left( \frac{%
\sin \xi e^{-\xi }}{\xi }\right)
\end{equation}
where $\xi =t_{F}/(\sqrt{2}\lambda _{J})$, $a_{0}$ is the lattice constant,
and we have reinserted the electric charge and Bohr magneton so that $a$ and
$b$ have units of a magnetic field. If we take $\lambda_{J}=20$ \AA , $%
a_{0}=2$\AA , $j_{e}= 10^{11}A/m^{2}$, we find $a=-1056$ (Oe) and $b=457$
(Oe) for a typical experiment with $t_{F}=25$\AA .

In conclusion we have found that by considering the exchange forces between
the conduction electron spin and the background magnetization for the spin
current perpendicular to the layers of a magnetic multilayer there exists
the effective field {\em and} torque, {\em both} of which contribute to
current driven reversal of the magnetization. We treat both terms on an
equal footing and demonstrate that they have a common origin. Our solution
differs in two important aspects from previous work: we find the {\em %
longitudinal} spin accumulation does not play a role in the switching, and
the spin torque, as well as the effective field, arises from a region 
in the magnetic layer within $\sim \lambda _{J}$ of the interface.
Therefore, the
decay length in our theory is related {\em neither} to the phase of the
wavefunction \cite{berger,slon}, {\em nor} to the spin diffusion length as
in the effective field concept of switching \cite{heide}.
We would like to
acknowledge our very fruitful conversations with Yaroslaw Bazaliy, Piet
Brouwer, Carsten Heide, Henri Jaffres, Barbara Jones, Roger Koch, Iouli
Nazarov, Dan Ralph, Andrei Ruckenstein and John Slonczewski. Many of them
took place this summer at the Aspen Center for Physics this summer and we
gratefully acknowledge its hospitality. This work was supported by the
National Science Foundation (DMR0076171), and the Defense Advanced Research
Projects Agency and Office of Naval Research (Grant No. N00014-96-1-1207 and
Contract No. MDA972-99-C-0009\ ).


\begin{references}
\bibitem{lambdaJ} This expression of $\lambda _{J}$ is strictly
valid when the momentum relaxation time $\tau $ is shorter than $
\hbar /J$. In the opposite limit $\lambda _{J}\approx v_{F}\hbar /J$; 
while this looks the same as a ballistic result, it comes from the
Boltzmann equation in the diffusive limit. As $\tau $ and $\hbar /J$
have the same order of magnitude in layers of ferromagnetic metals,
these two expressions for $\lambda _{J}$ lead to only slightly
different numerical values.
\bibitem{footnote}  To obtain a simple and transparent result we have
neglected the spin accumulation in the thick layer, i.e., we have assumed
the spin current in the thick layer is unchanged up to the interface. 
This is strictly valid only for half-metallic ferromagnetic layers. A more
realistic calculation should treat the spin accumulation in the thick layer
on an equal footing as the thin layer, i.e., in a self consistent manner;
also one should also include the the interface spin dependent scattering
(reflections and diffuse) which is omitted in our simple illustration.
\end{references}
\end{document}